\documentclass[aps,prd,nofootinbib,superscriptaddress,twocolumn]{revtex4-1}

\usepackage{latexsym}
\usepackage{amsthm}
\usepackage{amsmath}
\usepackage{graphicx}
\usepackage{amssymb}
\usepackage[hidelinks]{hyperref}

\usepackage{graphicx}
\usepackage{epstopdf}
\usepackage{amsmath}
\usepackage{amsfonts}
\usepackage{amssymb}
\usepackage{color}
\usepackage{mathrsfs}

\newcommand{\gsim}{\lower.7ex\hbox{$\;\stackrel{\textstyle>}{\sim}\;$}}
\newcommand{\lsim}{\lower.7ex\hbox{$\;\stackrel{\textstyle<}{\sim}\;$}}

\def\L{{\mathcal L}}
\def\mpl{M_{\rm Pl}}
\newcommand{\be}{\begin{equation}}
\newcommand{\ee}{\end{equation}}
\newcommand{\bea}{\begin{eqnarray}}
\newcommand{\eea}{\end{eqnarray}}

\newcommand{\comment}[1]{}
\newcommand{\expect}[1]{\left\langle #1 \right\rangle}
\newcommand{\mbf}[1]{\mathbf #1}
\newcommand{\mal}{\mathcal}

\def\ep{\epsilon}

\def\d{\partial}

\def \fnl{f_{NL}}

\def\x{\mbf x}
\def \k {\mbf k}
\def \O{\mal O}
\def\vep{\varepsilon}

\def\tsc{\theta_{\rm sc}}
\begin{document}

\pagestyle{plain}

\title{Gravitational Waves and the Scale of Inflation}

\author{Mehrdad Mirbabayi}

\affiliation{Institute for Advanced Study, Princeton, NJ 08540}

\author{Leonardo Senatore}

\affiliation{Department of Physics and SLAC,
Stanford University, Stanford, CA 94305, USA}

\author{Eva Silverstein}

\affiliation{Department of Physics and SLAC,
Stanford University, Stanford, CA 94305, USA}

\author{Matias Zaldarriaga}

\affiliation{Institute for Advanced Study, Princeton, NJ 08540}

\date{\today}
\begin{abstract}

We revisit alternative mechanisms of gravitational wave production during inflation and argue that they generically emit a non-negligible amount of scalar fluctuations. We find the scalar power is larger than the tensor power  by a factor of order $1/\ep^2$.   For an appreciable tensor contribution the associated scalar emission completely dominates the zero-point fluctuations of inflaton, resulting in a tensor-to-scalar ratio $r\sim \ep^2$.  A more quantitative result can be obtained if one further assumes that gravitational waves are emitted by localized sub-horizon processes, giving $r_{\rm max} \simeq 0.3 \ep^2$. However, $\ep$ is generally time dependent, and this result for $r$ depends on its instantaneous value during the production of the sources, rather than just its average value, somewhat relaxing constraints from the tilt $n_s$.
We calculate the scalar 3-point correlation function in the same class of models and show that non-Gaussianity cannot be made arbitrarily small, i.e. $\fnl \gtrsim 1$, independently of the value of $r$. Possible exceptions in multifield scenarios are discussed.

\end{abstract}

\maketitle

\section{Tensor emission during inflation}

Inflation stretches the vacuum fluctuations of graviton field to nearly scale-invariant super-horizon gravitational waves  
\be
\label{tvacuum}
\begin{split}
\expect{\gamma^s_{\mbf k}\gamma^{s'}_{\mbf k'}}_{\rm vac}&=(2\pi)^3\delta^{(3)}(\mbf k +\mbf k')\delta^{ss'} \frac{1}{2k^3} \mal P_{t,\rm vac},\\
\mal P_{t,\rm vac}& = \frac{2H^2}{\mpl^2}.
\end{split}
\ee
These tensor modes lead to $B$-type polarization of the CMB \cite{Zaldarriaga,Kamionkowski}. If experiments with sensitivity comparable to those currently taking data
were to make a conclusive detection (see e.g. \cite{Abazajian:2013vfg}), equation ({\ref{tvacuum}) would imply that the scale of inflation is  $H\simeq 10^{14}\rm GeV$. However, it has been suggested in~\cite{GW,Sorbo,Mukohyama}, that there are other (secondary) mechanisms of gravity-wave production during inflation which can dominate the primary effect \eqref{tvacuum}. This invalidates the above inference of the scale of inflation. In these examples the gravitons are not zero-point quantum fluctuations, and therefore a measurement of primordial $B$-modes would not be direct check of the quantization of gravity~\cite{Abazajian:2013vfg}. Given these prospects  now seems to be the right time to reexamine those mechanisms in more detail. 

The basic idea behind these secondary mechanisms is to assume there is a sector $X$ that constantly absorbs energy via its coupling to the inflaton field $\phi$ and emits gravitational waves. This emission can result from localized and nearly incoherent processes  that occur periodically. {This possibility can be motivated in field theory by assuming an approximate discrete shift symmetry~\cite{Behbahani:2011it}.} For instance, the $X$ sector can be composed of particles (strings) whose mass (tension) is a function of $\phi$, and  naturally arises in string theory motivated models of inflation where the inflaton is a monodromy-extended direction in field space \cite{Silverstein:2008sg}, but in a regime of parameters where a sector of the spectrum becomes light each time the field traverses an underlying circle of sub-Planckian period  (as in e.g. \cite{Trapped,Unwind}). As a concrete field-theory model consider 
\be
M_X^2 = M^2 \sin^2 \left(\frac{\phi}{f}\right),
\ee
with $\dot\phi_{\rm sr}/f\gg H$. Each time the mass goes through zero, there is a burst of particle production. If these massive particles subsequently decay or scatter each other, they will emit soft gravitons via Bremsstrahlung \cite{GW}. 

Another possibility is to have a process that acts coherently over a Hubble distance. The primary example is a $U(1)$ gauge field with a coupling to inflaton \cite{Barnaby,Sorbo}
\be
\label{FF}
\frac{\alpha}{f}\phi F\tilde F.
\ee
This causes a tachyonic instability of one of the two helicities of the gauge field. If the instability rate is faster than expansion rate (but not too faster to destabilize the inflation), a large helical field is generated and sources a polarized tensor field.  We refer to these mechanisms as ``coherent emission by extended configurations.''

The natural question to ask is whether these scenarios can compete with the zero-point fluctuations \eqref{tvacuum}. Obviously the energy density $\rho_X$ of the auxiliary sector must be a small fraction of total energy density of the universe $3\mpl^2 H^2$. The Friedmann equations imply
\be
\rho_X+p_X +\rho_\phi +p_\phi = -2\mpl^2 \dot H,
\ee
where $\rho_X+p_X\sim \rho_X$ is sourced by $\rho_\phi+p_\phi$ (the kinetic energy of inflaton in the slow-roll models). We therefore expect
\be
\label{rhoX}
\rho_X  \lesssim  \mpl^2 H^2\ep,
\ee
where $\ep \equiv  -\dot H/H^2$. 

The level of gravitational waves with frequency $\omega$ which can potentially be emitted by $\rho_X$ is roughly
\be
\gamma_{\omega} \sim \frac{\rho_X}{\mpl^2 \omega^2}.
\ee
Taking $\omega \sim H$ and using the upper bound \eqref{rhoX} yields 
\be
\gamma \lesssim  \ep,
\ee
which can still be much larger than $H/\mpl$ of vacuum fluctuations \eqref{tvacuum}. 

Because of the coupling to the inflaton field, which is the only source of energy during inflation, the $X$ sector necessarily emits scalar waves as well. The second natural question hence regards the level of scalar emission as compared to the vacuum fluctuations:
\be
\label{svacuum}
\begin{split}
\expect{\zeta_{\mbf k}\zeta_{\mbf k'}}_{\rm vac}=&(2\pi)^3\delta^{(3)}(\mbf k +\mbf k')\frac{1}{2k^3}\mal P_{s,\rm vac},\\
\mal P_{s,\rm vac} =&  \frac{H^2}{2\mpl^2\ep}.
\end{split}
\ee
Can the scalar emission be kept sub-dominant to $\mal P_{s,\rm vac}$? Even if not, can an observably large tensor to scalar ratio be explained by these mechanisms?

For the purpose of order of magnitude estimates and heuristic arguments, it is useful to characterize the primordial scalar and tensor power in terms of the number of quanta ($N_s$ and $N_t$) in each logarithmic interval of wavelength and in a volume of the same wavelength size. In the absence of emission (or absorption) these numbers remain conserved during the expansion of the universe. Since the fluctuations of vacuum and first excited level in a box are of the same order, these excitation numbers can be approximated by comparing the ratio of the actual power to the zero-point power: $N_t\sim \mal P_t/\mal P_{t,\rm vac}$ and $N_s\sim\mal P_s/\mal P_{s,\rm vac}$. Hence, in the presence of a secondary mechanism the tensor to scalar ratio $r = 4\mal P_t/\mal P_s$ would be modified from its usual value, $r=16\ep$, to\footnote{The inclusion of scalar sound-speed would lead to $\ep \to \ep c_s$ in all of our constraints and makes them stronger. See complementary discussions in~\cite{EFTofI,guido,senatoretalk,Baumann:2014cja} and \cite{EFTofI,senatoretalk,Creminelli} on the relation between $r$ and respectively the scalar and tensor sound-speed, in the absence of secondary emission.}
\be
\label{r}
r\sim 16 \ep \frac{N_t}{N_s}.
\ee

What allows us to make general statements about $N_t/N_s$, is the nearly exponential expansion of the universe:
\begin{itemize}

\item For any emission process that operates at a physical frequency $\omega$, there is only a short period of time of order $H^{-1}$ during which a given $k$-mode can be excited. 

\item Suppose this process transfers a total energy of $E$ per Hubble volume into gravity waves within a logarithmic interval of frequency around $\omega$. By the time the waves exit the horizon and freeze, the power dilutes by a factor of $\omega^{-4}$; a fact that follows from our definition 
\be
\label{omega4}
N_t \sim \frac{E}{\omega}\frac{H^3}{\omega^3}.
\ee
This strongly suppresses waves produced at $\omega\gg H$. In appendix~\ref{app:hard}, we will illustrate this point in a more concrete example by considering emission of hard gravitons in strong gravity regime. But in the main text concentrate on soft emission at $\omega \sim H$. \footnote{Both arguments also apply to the emission of scalar modes.}
\end{itemize}

\subsection{Incoherent emission by localized sub-horizon events}

The above observation leads to a dramatic simplification if the emission process happens deep inside the horizon: the details of the production mechanism does not matter anymore. The emission of Hubble wavelength tensor and scalar modes depends only on a few coarse-grained features of the process.

The $X$ sector can be thought of as effectively being composed of particles of mass $M$, with $N_X$ spontaneous tunneling events per Hubble time per Hubble volume. At each event a total energy $M$ is transferred from the time-dependent background field to $X$ sector. We assume different events are spatially out of phase with respect to one another, while temporally they can be correlated. In addition to its invariant mass, each $X$ particle is characterized by quadrupole and higher moments. The gravitational emission at Hubble wavelength due to the time variation of these moments is suppressed by powers of $lH$, where $l$ is the characteristic size of the $X$ particle.\footnote{More explicitly, the emission in this regime is dominated by the interaction ~\cite{Goldberger:2004jt}
\bea\nonumber
 &&\int d^4x  \int dt_X \gamma^{-1}\delta^4(x^\mu - x_X^\mu(\tau)) \; Q^{ij}(t_X) \; R_{0i0j}(x)\\ 
 &&= \int d\tau\;  Q^{ij}(t_X) \; R_{0i0j}(x_X)
\eea
where $\tau$ is the proper time of the particle emitting gravity waves, $R_{\mu\nu\rho\sigma}$ is the Riemann tensor, and $Q^{ij}$ is the intrinsic quadrupole of the particle. The standard formula of quadrupole emission gives (see e.g.~\cite{Weinberg})
\be
\frac{d E_t}{d\omega}\sim \frac{\omega^6 Q(\omega)^2 }{\mpl^2}\sim  \frac{\omega^4 Q(t)^2 }{\mpl^2}
\ee
where in the last step we have used that $Q(\omega)=\int dt \; e^{i\omega t} Q(t)\sim Q(t)/\omega$.  To relate this to the mass of the object, We write $Q\sim M l^2$, where $l$ is the typical size of the object. Taking $\omega\sim v/l$, where $v$($\ll c$) is the characteristic velocity, we obtain
\be
\frac{d E_t}{d\omega}\sim \frac{\frac{v^4}{l^4} M^2 l^4  }{\mpl^2}\sim v^4\,\frac{ M^2}{\mpl^2}\ .
\ee
\label{quad}} Large tensor emission requires relativistic and asymmetric processes.

Consider for instance the decay of a particle of mass $M$ into two relativistic jets. This process is accompanied by considerable emission of soft gravitons. In flat space, the energy radiated per unit solid angle per unit frequency can be calculated using the standard Bremsstrahlung formula (e.g. \cite{Weinberg})
\be\label{Brem}
\begin{split}
\left(\frac{dE_t}{d\Omega d\omega}\right)=&\frac{\omega^2}{2\pi^2M_P^2}
\sum_{N,M}\frac{\eta_N\eta_M}{(P_N\cdot k)(P_M\cdot k)}\\
&\times \left[(P_N\cdot P_M)^2-\frac{1}{2}m_N^2m_M^2\right].
\end{split}
\ee
Here the sum is over external particles, $\{P_N\}$ are their momenta, and $\eta=+1$ for in-going particles and $-1$ otherwise. The result for the decay process is
\be
\label{decay}
\frac{dE_t}{d\Omega d\omega}=\frac{1}{(2\pi)^3}\left(\frac{M}{2\mpl}\right)^2.
\ee
(This formula also serves as an order of magnitude estimate for soft graviton emission from other relativistic processes involving particles of mass $M$.) Comparing the flat spectral index $dE_t/d\omega \propto \omega^0$ to the dilution factor $\omega^{-3}$ of \eqref{omega4} implies that such a decay process during the inflation contributes mainly to the power of modes with $\omega\sim H$, as already anticipated.

Note also that the above formula is valid only asymptotically, that is, if the final states traverse distances much longer than the wavelength of gravitons. If they are caught by multiple subsequent scatterings, although there is emission from each scattering, the waves interfere coherently and destructively (Landau-Pomeranchuk effect). Since we are interested in emission at $\omega \sim H$ and since each mode spends roughly a Hubble time $H^{-1}$ at Hubble frequency, the net effect of multiple scattering is to suppress the power. (Different regimes of Bremsstrahlung from multiple scattering are reviewed in appendix \ref{LPM}.)

Hence we expect, in order to have the largest gravitational emission, each $X$ particle should participate in a single relativistic event. The exact amount of emition depends on details of the process. However among such events the decay \eqref{decay} seems to be the most efficient one. As long as different tunneling events are independent the total number of gravitons produced is obtained by summing over individual decays:
\be
\label{a_i/a_f}
N_t \sim \frac{n_X(t_f)}{H^3} \frac{d E_t}{d\omega}\sim N_X \left(\frac{M}{\mpl}\right)^2 \left(\frac{a(t_i)}{a(t_f)}\right)^3.
\ee
This implies a suppression factor if the lifetime of $X$ particles, $\Delta t= t_f - t_i$, is comparable or longer than the Hubble time. We henceforth assume the opposite regime.

Let us make two final remarks. First, if subgroups of $X$ particles emit coherently, for instance by $n$ of them merging into a bound-state which subsequently decays, \eqref{a_i/a_f} gets enhanced by a factor of $n$. We were unable to find a realistic model of this kind and leave this as an open possibility. 

Second, one should ask if the tunneling event itself leads to considerable gravitational emission. As already mentioned, if each event results in a localized massive object of small size (and hence small quadrupole moment), the emission is negligible. In more realistic scenarios particles are produced in pairs \cite{GW}. If these are two non-relativistic particles of mass $M/2$, that conclusion still holds. (In fact the most efficient emission process would then be for the two particles annihilating into two relativistic jets, so that the pair can effectively be treated as a single particle of mass $M$ all along.) On the other hand, if the pair is relativistic there is gravitational emission given exactly by \eqref{decay}. 

We postpone the actual calculation of de Sitter correlation function resulting from the decay process to appendix \ref{dS}.

\subsection{Coherent emission by extended configurations}

 As in the example of $U(1)$ gauge field, the tensor emission can happen coherently over horizon-size patches. Obtaining precise universal results seems impossible in this case. Nevertheless, an order of magnitude estimation of the maximum tensor emission, given energy $M$ per Hubble volume of an extended configuration, is easy. In the weak gravity regime we expect   
\be
\gamma(\omega\sim H) \sim \frac{M H}{\mpl^2}.
\ee
(As before, the main contribution comes from emission into nearly Hubble frequency modes.) We therefore get $\mal P_t \sim  M^2 H^2/\mpl^4$. If there are several extended configurations emitting independently (e.g. $N$ species of $U(1)$ gauge fields in the example of \cite{Mukohyama}), their contribution adds up to give
\be
\label{Ntext}
N_t \sim N \frac{M^2}{\mpl^2}.
\ee
Note that symmetries can highly suppress tensor emission compared to this naive expectation.

\section{Scalar emission and energy conservation}

Since the energy $M$ of each $X$ event is provided by the coupling of $X$ sector to the time-dependent inflaton background field, each tunneling event leads to scalar emission. By the same arguments as for gravitons [see \eqref{omega4}] only soft scalars of wavelength $\lambda\sim 1/H$ need to be considered. This emission can be calculated in a model-independent way using the Effective Field Theory of Inflation (EFTofI)~\cite{EFTofI,Senatore:2010wk}. However, to build intuition let us first consider a single-field slow-roll model. 

For deeply sub-horizon events curvature can be ignored and energy conservation implies that the energy $M$ must come from the background inflaton. Therefore, the $X$ particle production must be accompanied by a scalar wave ${\delta\phi}(t,x)$ which is responsible for that energy deficit. The stress-energy tensor of $\phi$ is
\be
T_{\mu\nu}= \d_\mu\phi \d_\nu \phi -g_{\mu\nu}\left[\frac{1}{2}(\d\phi)^2 -V(\phi)\right].
\ee
The perturbed energy density is
\be
\label{T00}
\delta T_{00} = \dot\phi_{\rm sr}{\delta\dot\phi} + V' {\delta\phi} + \O({\delta\phi}^2).
\ee
Terms quadratic in $\delta\phi$ correspond to the energy carried by the wave of ${\delta\phi}$, and are negligible compared to the linear terms.  Moreover, the scalar profile would be a sharp pulse with characteristic width determined by the size and duration of the tunneling event. As such most of the energy is carried away by frequencies higher than the Hubble rate.  Therefore, the second term of \eqref{T00} which can be written as $-3H\dot\phi_{\rm sr}{\delta\phi}$, is suppressed compared to the first one. By energy conservation $\delta T_{00}$ must integrate to $-M$, implying
\be
\label{phidot}
\int d^3x \; {\delta\dot\phi} = -\frac{M}{\dot\phi_{\rm sr}}.
\ee
(Note that the integral on the left is the Noether charge associated to the approximate shift symmetry of $\delta\phi$, whose mass is much less than $H$; it remains conserved after the tunneling event.)

Now consider the flat-space expression for the energy emitted in the wave of ${\delta\phi}$:
\be
E_s=\int d^3 x  \frac{1}{2}[{\delta\dot\phi}^2 + (\nabla {\delta\phi})^2].
\ee
Using \eqref{phidot} and dimensional analysis, we get at small frequencies 
\be
\frac{d E_s}{d\omega} \sim \frac{M^2}{\dot\phi_{\rm sr}^2}\omega^2.
\ee
 Because of the dilution effects associated with the expansion the scalar power would be most affected by the lower end of the spectrum, frequencies of order $H$. For these frequencies our flat space analysis is only an order of magnitude estimate.

\subsection{Generalization}

Consider an independent localized tunneling event in which energy $M$ is transferred to $X$ sector in a period much shorter than a Hubble time. By translational invariance of the background the total momentum transfer must be zero. As argued before for emission of long-wavelength tensor and scalars the detailed structure of the event is unimportant. Therefore we can describe it by the production of a particle at rest whose mass grows from $0$ to $M$. Choosing a frame where the constant-time hypersurfaces coincide with constant-inflaton hypersurfaces, the $X$ particle would be described by a DBI action with a time-dependent mass $M(t)$
\be
S_X= -\int d\tau M(t)\; \sqrt{g_{\mu\nu}\dot x_X^\mu \dot x_X^\nu}.
\ee
Following the EFTofI~\cite{EFTofI}, we can restore time-diffeomorphism invariance by shifting $t\to t+\pi$ and prescribing the right symmetry transformation to $\pi$.  Hence any explicit time-dependence [such as $M(t)$] leads to a linear coupling $\pi \d_t\L$. Since explicit time-dependence in the action results in energy non-conservation
\be
\d_\mu T^\mu_0= -\d_t \L,
\ee
on sub-horizon scales $\pi$ couples at leading order to $-\d_\mu T^\mu_0$. This coupling results in a universal emission of $\pi$ whenever energy is transferred from background to $X$ sector, and is a consequence of the conservation of total energy. Due to the mixing of $\pi$ and metric fluctuations $h_{\mu\nu}$ the gravitational stress-energy tensor of $\pi$ starts linear and is sign indefinite. Therefore, in any particle production event the total stress-energy tensor ($X$ plus $\pi$) is conserved well inside the horizon.

Up to slow-roll corrections the dynamics of scalar modes during inflation is adequately described by $\pi$ alone. { Restricting to the leading derivative operators in the EFTofI,} the relevant action for $\pi$ is therefore
\be
\label{Spi}
\begin{split}
& S =  -\int d^4x \sqrt{-g} \mpl^2\dot H [\dot \pi^2 - a^{-2}(\d\pi)^2] \\
&+ \int d^4x M(t+\pi) \int dt_X \gamma^{-1}\delta^4(x^\mu - x_X^\mu(\tau)),
\end{split}
\ee
where $\gamma = p_X^0/M_X$. For each independent event momentum conservation forces $\gamma =1$. The above action results in a cubic coupling between the canonically normalized field $\pi_c = \sqrt{2\ep}\mpl H\pi$ and $X$ particles with strength\footnote{The minimal scenario considered above (and in \cite{GW}) corresponds to $M=g\phi$ and $\delta\phi = \pi_c$.}
\be
\label{geff}
g_{\rm eff} \equiv \frac{\dot M}{\sqrt{2\ep}\mpl H }.
\ee
The relative factor of $\sqrt{-g}$ between the first and second line of \eqref{Spi} is responsible for a dilution effect similar to \eqref{a_i/a_f}. In the following we assume $H (t_f-t_i) \ll 1$.

The flat space solution for $\pi$ in the presence of a single source $X$ at $\x = 0$ is
\be
\begin{split}
\pi_c(t,\x) = &\frac{1}{\sqrt{2\ep}\mpl H}\int \frac{d^3\mbf k}{(2\pi)^3} e^{i\k\cdot \x}\\
&\left[\frac{i e^{-i k t}}{2k}\int_{t_i}^{t_f}\dot M e^{ik \tau} dt' + \rm{c.c.}\right],
\end{split}
\ee
which after integration by parts gives
\be
\label{pic}
\pi_c(t,\x) = \frac{M(t_f)}{\sqrt{2\ep} \mpl H}\int \frac{d^3\k}{(2\pi)^3 k} e^{i\k\cdot \x} \sin k(t-t_f),
\ee
plus terms suppressed by $\O(k(t_f-t_i))$ which is negligible because we are interested in $k\sim H$. Using the expression for the energy density of the canonically normalized field in flat space: $(\dot \pi_c^2 +(\nabla\pi_c)^2)/2$ we obtain the following result for the energy emitted per unit frequency per unit solid angle
\be
\label{Es}
\frac{dE_s}{d\omega d\Omega}= \frac{1}{(2\pi)^3}\frac{M^2\omega^2}{4\ep \mpl^2 H^2}.
\ee
As before, the total scalar emission is obtained by summing over independent $X$ events. The maximum ratio $N_t/N_s$ can therefore be calculated by comparing \eqref{decay}, which {appears to be} to the most efficient gravity wave production scenario, with the above formula at $\omega \sim H$. This yields\footnote{We note that the $\ep$ appearing here is the instantaneous $\ep$ at the time of $X$ production. It is natural to expect $\ep$ to have periodic variations in the inflationary models with particle production, however the relative amplitude of oscillations is small in the conventional models \cite{Behbahani:2011it}. Models with significant variations in $\ep$ can perhaps be constructed by considering non-monotonic inflationary potentials. In this case one of the $\ep$ factors in the bound \eqref{rmax} has to be replaced by the instantaneous value of $\ep$ at scalar emission.}
\be
\label{Nt/Ns}
\frac{N_{t,\rm max}}{N_s}{ \sim \ep}.
\ee
Evidently, in order for the production mechanism to have any significance, that is $N_t\sim 1$ or larger, the scalar byproducts would completely dominate the vacuum fluctuations \eqref{svacuum}. Therefore according to \eqref{r}, the largest possible tensor to scalar ratio which can be obtained {in this scenario} is of order $\ep^2$. The more careful calculation of scalar and tensor correlation functions in de Sitter space (appendix \ref{dS}) yields
\be
\label{rmax}
r_{\rm max} \simeq 0.3 \ep^2.
\ee
An observable level of $B$-modes, {$r\gtrsim 10^{-3}$}, requires $\ep > { 0.05}$ in these scenarios. The Hubble parameter would therefore drop by about one order of magnitude during the $60$ e-folds of inflation. Nevertheless, the scalar and tensor tilt can remain sufficiently small as we will estimate in the next section. From this estimate a value of $r\gtrsim 0.1$ seems hard to be explained with this class of models, although that may depend on the extent to which $\ep$ varies.\footnote{In the example of tensor emission by long string pairs considered in \cite{GW}, the scalar emission during the process of energy transfer from background to the $X$ sector was missed. This led to a much larger estimates for $N_t/N_s$. In fact the tensor to scalar ratio is much smaller in this case ($r\sim \ep^2/N_{\rm loop}$), because the initial scalar emission by the long string pair is coherent, while the subsequent gravitational emission by the decay of the pair into $N_{\rm loop}$ pieces is incoherent.}

The scalar emission by extended objects can be estimated as follows. The $\pi\d_\mu T^\mu_0$ coupling and the overall normalization of $\pi$ kinetic term in \eqref{Spi} imply 
\be
\pi(\omega\sim H)\sim \frac{M H^2}{\mpl^2 \dot H} ,
\ee
for energy $M$ of the configuration in a Hubble volume. Using $\zeta \simeq -H\pi$ gives $\mal P_s\sim H^2 M^2/\mpl^4 \ep^2$. Comparison with $\mal P_{s,\rm vac}$ gives
\be
N_s\sim N\frac{M^2}{\ep \mpl^2}
\ee
where we also inserted the number of species $N$. Comparing with \eqref{Ntext} gives $r_{\rm max}\sim \ep^2$.

\subsection{Multifield inflation}

In the context of tensor emission by $U(1)$ gauge field production, it has been suggested \cite{Mukohyama} to decouple scalar emission by introducing another scalar field $\psi$ which is slow-rolling (say with $\dot \psi_{\rm sr}\ll \dot\phi_{\rm sr}$):
\be
\begin{split}
S=&\int \sqrt{-g}\left[\frac{\mpl^2}{2} R +\frac{1}{2}(\d\phi)^2 +\frac{1}{2}(\d\psi)^2\right. \\
& \left.  -V(\phi,\psi) -\frac{1}{4} F^2 -\frac{\psi}{4f}F \tilde F \right].
\end{split}
\ee
By replacing \eqref{FF} with $\frac{1}{f}\psi F\tilde F$ the energy source of the auxiliary sector becomes $\dot\psi_{\rm sr}^2/2$. If in addition the scalar spectrum is exclusively determined by $\phi$ fluctuations, then energy conservation doesn't seem to enforce any correlation between tensor and scalar emission. Note that this idea, if viable, can be used in other production mechanisms as well. But is it viable?

Let us first understand this proposal in view of the above argument for the universal $\pi$ emission as a consequence of energy conservation. In the EFTofI the fluctuations of fields are decomposed into parallel (adiabatic) and perpendicular (iso-curvature) to the background trajectory in field space~\cite{Senatore:2010wk}. The $\pi$ field, which we refer to as the inflaton, is the fluctuations along the background trajectory. As such, it is a linear combination of the fluctuations of $\phi$ and $\psi$, and after canonical normalization reads
\be
\label{pic2}
\pi_c =\frac{1}{\sqrt{\dot\phi_{\rm sr}^2+\dot\psi_{\rm sr}^2}}(\dot\phi_{\rm sr}\delta\phi +\dot\psi_{\rm sr}\delta\psi).
\ee
In particular it couples to the gauge fields:
\be
\label{piA}
\mal{L}_{\pi A}=\frac{\alpha}{4f}\pi_c F\tilde F, \quad \alpha \simeq \frac{\dot\psi_{\rm sr} }{\dot\phi_{\rm sr}}.
\ee
There is also a light field $\sigma$ which characterizes perpendicular fluctuation in field space, and has a similar coupling to $F\tilde F$ but with $\alpha \sim 1$. 

The copious production of gauge fields excites $\pi$ and results in a contribution to the scalar power that is the same as the single-field version of the model \cite{Barnaby}
\be
\mal P_s = \mal P_{s,\rm vac}\left[1+7.5\times 10^{-5} \mal P_{s,\rm vac}\frac{e^{4\pi \xi}}{\xi^6}\right],
\ee
with $\xi = \alpha \dot\phi_{\rm sr}/2 f H$. Comparing to the tensor production \cite{Mukohyama}
\be
\label{gaugetensor}
\mal P_t = 16 \ep \mal P_{s,\rm vac}\left[1+3.4\times 10^{-5} \ep \mal P_{s,\rm vac}\frac{e^{4\pi \xi}}{\xi^6}\right],
\ee
we see the same $\ep^2$ suppression. As expected $\pi$ couples to $\d_\mu T^\mu_0$ and whenever there is particle production (gauge fields in this case) there is an associated emission of $\pi$.\footnote{ We notice in passing that the signal in this model is exponentially sensitive to the value of $\xi$, so that it is observationally relevant only for a very small range of values of~$\xi$. See \cite{Ozsoy,Ferreira} for a more detailed study of this model, and \cite{Matteo} for a related work on multi-field scenarios.}

However, the iso-curvature fluctuations ($\sigma$) in this model are also sourced by the gauge fields. The resulting iso-curvature modes can later convert into adiabatic ones~\cite{Senatore:2010wk}:
\be
\label{zetasigma}
\zeta \simeq -H\pi +\zeta_{,\sigma}\sigma.
\ee
Therefore, if $\zeta_{,\sigma}\simeq H\dot\psi_{\rm sr}/\dot\phi_{\rm sr}^2$ the net contribution of the gauge fields to observed scalar spectrum can be made negligible. 

Since this conversion happens at super-horizon scales, it is indeed easier to work in terms of the background model formulated in terms of $\phi$ and $\psi$, and ask if it is possible to decouple $\zeta$ from $\delta\psi$, up to possible slow-roll suppressed corrections. This problem is studied in more detail in appendix \ref{multi}.  It is argued that choosing the reheating surface to be determined by $\phi$ and $\ep \ll 1/N_e$ (where $N_e\sim 60$ is the number of e-folds of inflation), decouples $\delta\psi$ and corresponds to the above value for $\zeta_{,\sigma}$. 

Another multifield inflationary model that evades our conclusion because of a fundamentally different reason is chromonatural inflation \cite{Adshead}. Here there is a non-abelian gauge field background which causes the perturbations of the gauge field mix with the tensor modes. The assumption that tensor modes couple universally via a cubic coupling of strength $1/\mpl$ does not hold in this example since the gauge field fluctuations can directly oscillate into gravitons.

\section{Non-Gaussianity, and tilt \label{ng}}

In the last section we argued that for a large class of models, large gravitational emission implies dominant scalar emission. The scalar spectrum is naturally expected to be non-Gaussian. In the case of localized emission, the non-Gaussianity is calculated in appendix \ref{dS}, and in terms of the conventional $\fnl$ parameter 
\be
\label{fnlz}
\fnl \zeta \sim \frac{\expect{\zeta^3}}{\expect{\zeta^2}^{3/2}}\sim N_X^{-1/2},
\ee
where $N_X\sim n_X H^{-3}$, the number of $X$ particles per Hubble volume, is related to $N_s$ through \eqref{Es} and the assumption of incoherent emission, namely
\be
\label{Ns}
N_s \sim N_X \frac{M^2}{\ep \mpl^2}.
\ee
In \eqref{fnlz}, $\zeta$ is a shorthand:
\be
\label{zeta}
\zeta \equiv \mal P_s^{1/2}= N_s^{1/2} \mal P_{s,\rm vac}^{1/2} \sim 10^{-5}.
\ee
By energy conservation $N_X$ cannot be made arbitrarily large, therefore there is a lower bound on $\fnl$.\footnote{Another possibility is to have yet another contribution to the scalar spectrum ($N_s'$) with a Gaussian distribution and $N_s'\gg N_s$. This is unlikely to allow any observable gravity wave signal as it suppresses the ratio \eqref{rmax} by another factor of $N_s/N_s'$.} Let us rewrite the constraint \eqref{rhoX} using the average number density $n_X$ of tunneling events and the final mass $M$:
\be
\label{rhoX2}
\rho_X = n_X M \lesssim \mpl^2 H^2 \ep.
\ee
Multiplying both sides by a Hubble volume $1/H^3$ gives
\be
\label{NsH}
N_X \frac{MH}{\ep \mpl^2} \lesssim  1.
\ee
Using \eqref{Ns} the l.h.s. can be written as $N_X^{1/2}N_s^{1/2}H/\mpl \sqrt{\ep}$, which together with \eqref{fnlz} and \eqref{zeta} results in
\be
\label{fnl}
\fnl  \gtrsim 1.
\ee
The large non-Gaussianity can be used to break the degeneracy between the production scenarios and the conventional one. In Appendix~\ref{dS} we provide the shape of the induced non-Gaussianity. We find that the cosine with the standard equilateral~\cite{Creminelli:2005hu} and orthogonal~\cite{Senatore:2009gt} templates is quite large: 0.97, making it rather challenging to actually distinguish this shape from the standard ones.

We should emphasize that the above degree of non-Gaussianity \eqref{fnlz}, which arises from stochasticity of the emission process, is the minimum level based on very general assumptions. In any concrete model there may be other sources of non-Gaussianity--which we do not expect them to cancel the stochastic piece. For instance, if after tunneling the $X$ particles remain coupled to inflaton for a longer period of time, their scalar emission would be influenced, and hence correlated, with the previously emitted waves. This is the case in the ``Trapped Inflation'' model \cite{Trapped}. The already existing non-Gaussianity bounds may then lead to more stringent constraints on the maximum level of gravitational wave emission as argued in \cite{Ozsoy}. Note however that various sources of non-Gaussianity are expected to be indep


{\bf Tilt:} We finally calculate the tilt of scalar and tensor spectra assuming that the coupling strength $g$ and mass $M$ are dictated by UV physics and remain approximately constant during inflation. Since $\dot M$ is the only relevant dimensionful parameter at the time of tunneling, the number density of $X$ events is expected to be $n_X\sim \dot M^{3/2}$ or $N_X\sim (\dot M H^2)^{3/2}$. Using the formulae \eqref{geff}, \eqref{zeta}, and \eqref{Ns}, we get
\be
\zeta^2 \sim g^{3/2}\frac{M^2}{\mpl^2} \left(\frac{H}{\mpl}\right)^{1/2}\ep^{-5/4}.
\ee
The scalar tilt is therefore
\be
n_s -1 =-\frac{1}{2}\ep -\frac{5}{4}\ep_2,
\ee
where $\ep_2 \equiv \dot \ep /\ep H$. A similar calculation gives for the tensor tilt
\be
n_t = -\frac{1}{2}\ep +\frac{3}{4}\ep_2.
\ee
Thus, even $\ep \simeq 0.1$ does not require significant fine-tuning. Here we did not make a distinction between the instantaneous and average $\ep$, but that only relaxes the constraints from $n_s$.
\section{Conclusions}

{We have analyzed a large class of inflationary models, and found that} there is a generic upper bound on the tensor to scalar ratio in secondary mechanisms of tensor production. This is because emitting gravitons requires energy, and transferring energy from time-dependent background to another sector leads to coupling to inflaton $\pi$ and scalar emission. This emission is expected to be larger than the tensor emission by a factor of $1/\ep^2$. Therefore, unless there are iso-curvature modes which are produced by the same process and cancel the former contribution to $\zeta$, tensor-to-scalar ratio will be $\O(\ep^2)$. These models are also associated with large non-Gaussianity, i.e. $f_{NL}\gtrsim 1$, independently of the value of $r$. In the event of a detection of $r$, non-Gaussianities would provide a way to potentially, though challengingly, distinguish this scenarios from the standard one.

\section*{Acknowledgments}

We thank S.~Adler, G.~D'Amico, R.~Flauger, A.~Gruzinov, M.~Kleban, S.~Mukohyama, R.~Namba, M.~Peloso, and G.~Shiu for stimulating discussions. M.M. is supported by NSF Grants No. PHY-1314311 and No. PHY-0855425. L.S.~is supported by by DOE Early Career Award DE-FG02-12ER41854 and the NSF Grant PHY-1068380.    E.S. is supported  in part by the National Science Foundation
under grant PHY-0756174 and NSF PHY11-25915 and by the Department of Energy under
contract DE-AC03-76SF00515. M.Z. is supported in part by the NSF grants AST- 0907969, PHY-1213563 and AST-1409709.

\appendix

\section{Hard gravitons\label{app:hard}}

Only well-localized objects can emit hard gravitons. Clearly the effect is larger in the strong gravity regime. So consider the production of $n_X$ mass-$M$ black hole pairs per unit volume which subsequently merge and emit an order one fraction of their mass into gravitons of frequency 
\be
\omega \sim \frac{1}{r_g}\sim \frac{\mpl^2}{M}.
\ee
The number of tensor modes is of order
\be
N_t\sim \frac{n_X M}{\omega^4}\sim \frac{n_X}{\omega^3}\frac{M^2}{\mpl^2},
\ee
where $n_X\ll \omega^3$ but $M\gg \mpl$, so there is a chance of having $N_t>1$. However, when compared to the associated soft scalar emission:
\be
N_s\sim \frac{n_X}{H^3} \frac{M^2}{\ep \mpl^2}
\ee
one obtains the much smaller tensor to scalar ratio
\be
r \sim \ep^2 \frac{H^3}{\omega^3}.
\ee

\section{de Sitter correlators\label{dS}}

To calculate de Sitter correlation functions of $\pi$ one has to take the expansion of the universe and periodicity of the production mechanism into account. Suppose there is a sequence of random production of $X$ particles with average proper density $\bar n_X$ and at moments $\eta_n$. Let's at each $\eta_n$ divide the space into small cells $i$, each of comoving volume $\delta v_i$ so small that $p_i\equiv \bar n_X a^3 \delta v_i\ll 1$. To each cell assign a random variable $X_{i,n} = 1$ with probability $p$, and $0$ otherwise. As we saw in \eqref{pic} each event is practically a delta function source for $\pi$. Hence, the field $\pi_{\k}(\eta)$ resulting from creation events can be written
\be
\pi_\k(\eta) =\frac{M}{2\ep\mpl^2 H^2} \sum_n G_\k(\eta,\eta_n)\sum_i X_{i,n} e^{i\k\cdot \x_i}
\ee
where the first sum is over the production times and the second on the cells. The de Sitter retarded Green's function in the limit $k\eta \to 0$ simplifies to 
\be
\label{g}
G_\k(0,\eta_n)= \frac{H^2}{k^3}(\sin k\eta_n - k\eta_n \cos k\eta_n) \equiv \frac{H^2}{k^3}g(k\eta_n).
\ee
The late-time 2-point function of $\pi$ can be calculated by noting that
\be
\expect{X_{i,n} X_{j,m}} = \bar n_X^2 a_n^2 a_m^2 \delta v_i \delta v_j + \bar n_X a_n^3\delta v_i \delta_{nm}\delta_{ij},
\ee
The first term gives a disconnected contribution proportional to $\delta^3(\k_1)\delta^3(\k_2)$, but the second term gives
\be
\label{2}
\expect{\pi_{\k_1}\pi_{\k_2}}'=
\frac{M^2}{(2\ep\mpl^2)^2 k^3} 
\left[\frac{\bar n_X}{H^3} \sum_{n} \frac{g^2(k\eta_n)}{-k^3\eta_n^3}\right].
\ee
Where prime means that $(2\pi)^3 \delta^3(\k_1 +\k_2)$ has been omitted. The sum gets contribution only from those $\eta_n$ for which $k\eta_n = \O(1)$. Thus the expression in brackets is $\O(N_X)$. The calculation of 3-point function of $\pi$ is very similar and gives
\be
\label{3}
\expect{\pi_{\k_1}\pi_{\k_2}\pi_{\k_3}}'= \frac{M^3}{(2\ep\mpl^2)^3 k_1^2 k_2^2 k_3^2}\left[\frac{\bar n_X}{H^3} \sum_n 
\prod_i \frac{g(k_i\eta_n)}{-k_i\eta_n}\right].
\ee
The expression in the bracket is $\O(N_X)$. From \eqref{g} it follows that in the squeezed limit $k_1\to 0$ the above expression scales as $\O(k_1^0)$. The self-interactions of $\pi$ and slow-roll suppressed higher order terms in the conversion between $\pi$ and $\zeta$ will change the squeezed limit behavior of the 3-point function. 

{\bf Tensors:} The calculation of tensor power is similar. One first defines the helicity components of the transverse-traceless part of the spatial metric
\be
\gamma_{\k, ij} = \gamma^+_\k \vep^{+}_{ij}+\gamma^-_\k \vep^{-}_{ij},
\ee
where $\vep^{\pm}_{ij}$ are $\k$-dependent polarization tensors: $\vep^{\pm}_{ij}= \vep^{\pm}_i\vep^{\pm}_j/\sqrt{2}$, with $ {\vep^{\pm}} = {\hat x} \pm i {\hat y}$ for $\k \propto {\hat z}$. The linear equation of motion for the helicity modes is
\be
{\gamma^r_\k}'' -\frac{2}{\eta}{\gamma^r_\k}'+k^2 \gamma^r_\k 
= 16\pi G a^2 \frac{1}{2\sqrt{2}}\vep^r_i \vep^r_j T^{ij}_\k.
\ee
In the decay of a particle of mass $M$ into two relativistic jets the r.h.s. gets contribution only from the stress tensor of the jets:
\be
T^{ij}=\frac{\eta}{\eta_n} E {\hat p}^i {\hat p}^j 
a^{-3}\delta^{(3)}(\x - {\hat p} (\eta-\eta_n)) \theta(\eta-\eta_n)
\ee
where $\eta_n$ is the decay time and ${ p}=E {\hat p}$ the momentum of the jet. For an event at $\x =0$ one obtains (after summing the contribution of the two jets)
\be
\gamma^r_\k(\eta) =\frac{M}{\sqrt{2}\mpl^2}({\hat\vep^r}\cdot {\hat p})^2
\frac{1}{H\eta_n}\int_{\eta_n}^0 d\eta' G_\k (\eta,\eta')
\cos ({\hat p\cdot k} \eta') .
\ee
Hence, to calculate the emission from multiple events one needs to keep track of ${\hat p}$ of each event in addition to its position $\x_i$ and time $\eta_n$. We therefore define the random variable $X_{i,\hat p,n} = 1$ with probability $p_i=n_Xa_n^3 \delta v_i d^2\hat p/4\pi$, and $0$ otherwise. A similar calculation as in the scalar case leads to
\be
\label{2t}
\begin{split}
\expect{\gamma^r_\k \gamma^s_{\k'}}'= \frac{\delta^{rs}}{2k^3}P_{t,\rm vac}
\frac{M^2}{\mpl^2}
\left[\frac{\bar n_X}{H^3}\sum_n \frac{1}{-2k^3\eta_n^3}\right. \\
\left.\int_0^1 d\mu \left(\frac{1-\mu^2}{\eta_n}
\int_{\eta_n}^0 d\eta g(k\eta) \cos \mu k\eta\right)^2\right].
\end{split}
\ee
Comparing to \eqref{2} yields
\be
\label{Nt/NsdS}
\frac{\mal P_t/\mal P_{t,\rm vac}}{\mal P_s/\mal P_{s,\rm vac}} = \ep \frac{\sum_n H(k\eta_n)}{\sum_n G(k\eta_n)}
\ee
where $G$ and $H$ represent the summands in \eqref{2} and \eqref{2t}~\footnote{ For $-k\eta_n\gg 1$, $G\simeq -\cos^2 k\eta_n/k\eta_n$ while $H\simeq (\cos 2 k\eta_n-2)/12 k^3\eta_n^3$.  Averaging over several periods $1/k\ll \Delta\eta\ll -\eta_n$ we get for the ratio
\be
\frac{\expect{G(k\eta_n)}_{\Delta\eta}}{k^2\eta_n^2 \expect{H(k\eta_n)}_{\Delta\eta}}\simeq 3,
\ee
which is in rough agreement with the ratio $2$ inferred from the flat space results \eqref{decay} and \eqref{Es} (recall that \eqref{decay} is the energy emitted in both tensor polarizations). }

To get a more accurate result, we assume uniform distribution of creation events in physical time, and approximate the sums in \eqref{Nt/NsdS} by the integral $\int \frac{dt}{\Delta t}=\int_{-\infty}^0 d\eta_n/(\eta_n \Delta tH)$, with $\Delta t$ being the time-spacing of the events,  giving $\frac{\mal P_t/\mal P_{t,\rm vac}}{\mal P_s/\mal P_{s,\rm vac}}\simeq 0.018 \ep$ or
\be
r_{\rm max} \simeq 0.3 \ep^2.
\ee

\section{Bremsstrahlung emission from multiple scatterings\label{LPM}}

In this appendix we briefly review different regimes of soft emission in multiple scattering processes following \cite{Klein,Koshelkin}. The original derivation is for radiation by a relativistic electron moving inside matter. The Bremssrahlung emission of soft photons now depends on the rate of scattering $\Gamma$ and the average scattering angle per unit length $q=\expect{\theta_{\rm sc}^2}/l$. Replacing $\alpha \to M^2/\mpl^2$, where $M$ is the center of mass energy of the scattering process, gives a rough estimate of gravitational Bremsstrahlung. There are three regimes:

{\bf (I) Small angle:} Consider a single scattering event with a very small angle $\tsc \ll \gamma^{-1}$. The coherence time $\tau_1$ over which the electron can influence the emitted photon is
\be
\tau_1 (1-v)\sim \lambda.
\ee
Using $1-v \sim \gamma ^{-2}$ we get
\be
\label{tau1}
\tau_1\sim \frac{\gamma^2}{\omega}\gg \frac{1}{\omega}.
\ee
Since we are interested in gravitational waves of Hubble wavelength emitted in a Hubble time, we should consider emission during 
\be
T\sim \frac{1}{\omega}  \sim \frac{1}{H}.
\ee
Since $\tau_1\gg T$, the whole process can be approximated by a single event in this regime. Moreover, the energy emitted per unit frequency receives a small angle suppression
\be
\label{I}
\frac{d E}{d\omega}\sim \alpha\frac{\Delta p^2}{m^2}\sim \alpha \gamma^2 \tsc^2  \ll \alpha.
\ee
The above treatment is valid even if there are multiple scatterings, as long as $q\tau_1\ll \gamma^{-2}$.

{\bf (II) Landau-Pomeranchuk:} Suppose $q$ is increased beyond that. The new coherence length $\tau_2$ becomes shorter and will be determined in terms of $\theta_{\rm sc}=\sqrt{q\tau_2}>\gamma^{-1}$ according to
\be
\tau_2 c(1-\cos \theta_{\rm sc})\sim \lambda,
\ee
or
\be
\tau_2\sim \frac{1}{\omega\tsc^2}\gg \frac{1}{\omega},
\ee
where we assumed $\theta_{\rm sc}$ is still much less than unity. This implies
\be
\tau_2 = \frac{1}{\sqrt{\omega q}}\quad \text{and}\quad \tsc^2 =\sqrt{\frac{q}{\omega}}.
\ee
The requirement $\gamma^{-1}\ll \tsc \ll 1$ in one coherence length $\tau_2$ gives
\be
1\ll \frac{\omega}{q}\ll \gamma^4.
\ee
The emission rate can be computed as follows. One first cuts the particle trajectory into coherent pieces of length $\tau_2$. Each segment can be thought of as a particle moving in a straight line whose charge is turned on at some moment and off after $\tau_2$. The emission from different segments add up incoherently. Unlike \eqref{I}, there is no small angle suppression since $\tsc \gamma \gg 1$, therefore:
\be
\label{II}
\frac{d E}{d\omega}\sim \frac{T}{\tau_2}\alpha=T \alpha\sqrt{q\omega}.
\ee
This is the standard Landau-Pomeranchuk formula. Note that for $T = 1/\omega$ this is a suppression compared to a single large angle scattering, giving
\be 
\frac{d E}{d\omega}\sim \alpha\sqrt{\frac{q}{\omega}}\ll\alpha.
\ee

\vskip 0.5 cm

{\bf (III) Large angle:} Finally, when $\omega \ll q$ there will be a lot of scatterings of order-one angle in a wavelength. Now the emission can be obtained by dividing the electron trajectory into segments of length $\tau_3=1/q$. Each segment makes an order one angle with the next one and hence emits incoherently. However, now the segment is shorter than the wavelength which results in a suppression of $\omega \tau_3$ in the amplitude. The emission formula becomes
\be
\label{III}
\frac{d E}{d\omega}\sim {Tq} \alpha \frac{\omega^2}{q^2}.
\ee

We conclude that the maximum amount of Bremsstrahlung emission during $T\sim 1/\omega$ is obtained by a single large angle scattering which results in $dE/d\omega\sim \alpha$.

\section{Secondary gravitational emission in Multifield inflation\label{multi}}

Suppose there are two scalar fields $\phi$ and $\psi$ rolling during inflation, and the energy for the auxiliary sector responsible for the gravitational emission is provided by coupling to $\psi$. The energy transfer from the background $\dot\psi$ to this sector leads to emission of $\delta\psi$ quanta. The contribution of these fluctuations to adiabatic modes can be obtained using $\delta N$ formalism \cite{Sasaki}, where $\zeta=\delta N$ is calculated by determining how much the expansion,
\be
\label{N}
N=\int dt H(t),
\ee
differs in different patches of the universe with different values of $\psi$ (and other fields). If eventually the $\psi$ fluctuations source adiabatic modes with proportionality coefficient
\be
\label{Npsi}
N_{,\psi} \sim \frac{H\dot\psi_{\rm sr}}{\dot\phi_{\rm sr}^2},
\ee
then the $r< \ep^2$ bound still remains in place. (Comma denotes partial derivative.) This is because for $\dot\psi_{\rm sr}\ll \dot\phi_{\rm sr}$ 
\be
\pi\simeq \frac{\pi_c}{\dot\phi_{\rm sr}} \simeq \frac{\delta \phi}{\dot\phi_{\rm sr}}
+\frac{\dot\psi_{\rm sr}}{\dot\phi_{\rm sr}^2}\delta\psi,
\ee
hence the contribution \eqref{Npsi} of $\delta\psi$ to $\zeta$ would be of the same order of magnitude as if $\zeta_{,\sigma}$ were absent from \eqref{zetasigma}.

Let us see when a contribution of order \eqref{Npsi} should be expected. The slow-roll condition $3H\dot\psi_{\rm sr} \simeq -V_{,\psi}$ implies that fluctuations in $\psi$ contain energy $\delta \rho_\psi \sim H\dot\psi_{\rm sr} \delta\psi$. Therefore, if these fluctuations perturb the kinetic energy of the inflaton $\dot\phi_{\rm sr}^2/2$ they result in \eqref{Npsi}. This will be the case for instance if $\delta\psi$ fluctuations become massive as function of $\phi$ and before the end of inflation since as in the examples studied in the text the available source of energy during inflation is the kinetic energy of the inflaton. Immediately after the transition we expect $\delta \dot\phi^2\sim \delta\rho_\psi$. On the other hand if the $\phi$ and $\psi$ sectors are two completely decoupled slow-rolling sectors, we expect
\be
N_{,\psi}\sim \frac{U(\psi)}{V(\phi)} \frac{H}{\dot\psi}.
\ee
This is $\ep_\phi/\ep_\psi$ times \eqref{Npsi}, where $\ep_\phi = \dot\phi_{\rm sr}^2/2 V(\phi)$ and similarly $\ep_\psi=\dot\psi_{\rm sr}^2/2 U(\psi)$. This would lead to $r_{\rm max} \sim \ep_\psi^2$. For $\psi$ to slowly roll $\ep_\psi$ must be small though perhaps it can be larger than $\ep$.\footnote{This scenario has been studied in more detail in \cite{Ferreira}. To compare note that their $\Delta N$ would be of the order of our $1/\ep_\psi$. In particular, $\Delta N$ should be long enough for the cosmologically relevant range of modes to cross the horizon.}

Therefore, to avoid \eqref{Npsi} one way is to couple the two sectors in such a way that the fractional contribution of $\delta \psi$ to expansion history be given by $\delta\rho_\psi/\rho_{\rm tot}$. This would be the case if the reheating surface is completely determined by $\phi$ independently of the value of $\psi$. As a toy model consider a potential
\be
U(\phi,\psi) = \theta(\phi_0 - \phi)V(\phi,\psi),
\ee
where $V(\phi,\psi)$ satisfies slow-roll condition for both fields. At $\phi=\phi_0$ all of the potential energy abruptly converts into the kinetic energy of $\phi$ at $\phi_0$. The fluctuations of $\psi$ induce fluctuations in $\dot\phi$ but since now $\dot\phi^2/2$ contains most of the energy density of the universe the fractional variations do not have the previous $1/\ep$ enhancement.

The fluctuations $\delta\rho_\psi/\rho_{\rm tot}$ also lead to variations in the number of e-folds \eqref{N} which can be approximated as:
\be
N\simeq\frac{1}{\mpl^2} \int^{\phi_0} \frac{V}{V_{,\phi}}d\phi.
\ee
For concreteness suppose $V_{,\phi,\psi}=0$. Then
\be
\label{NpsiNe}
N_{,\psi}\sim N_e \frac{V_{,\psi}}{V}\sim N_e \ep \frac{ H\dot\psi_{\rm sr}}{\dot\phi_{\rm sr}^2},
\ee
where $N_e\sim 60$ is the total number of e-folds from the horizon crossing of $\delta\psi$ fluctuations until the end of inflation. Hence if $N_e \ep\ll 1 $ the contribution of $\delta\psi$ fluctuations to scalar power can be suppressed.



\begin{thebibliography}{10}

\bibitem{Zaldarriaga} 
  M.~Zaldarriaga and U.~Seljak,
  ``An all sky analysis of polarization in the microwave background,''
  Phys.\ Rev.\ D {\bf 55}, 1830 (1997)
  [astro-ph/9609170].

\bibitem{Kamionkowski} 
  M.~Kamionkowski, A.~Kosowsky and A.~Stebbins,
  ``Statistics of cosmic microwave background polarization,''
  Phys.\ Rev.\ D {\bf 55}, 7368 (1997)
  [astro-ph/9611125].

\bibitem{Abazajian:2013vfg}
  K.~N.~Abazajian, K.~Arnold, J.~Austermann, B.~A.~Benson, C.~Bischoff, J.~Bock, J.~R.~Bond and J.~Borrill {\it et al.},
  ``Inflation Physics from the Cosmic Microwave Background and Large Scale Structure,''
  arXiv:1309.5381 [astro-ph.CO].

%
%

\comment{
\bibitem{Lyth} 
  D.~H.~Lyth,
  ``What would we learn by detecting a gravitational wave signal in the cosmic microwave background anisotropy?,''
  Phys.\ Rev.\ Lett.\  {\bf 78}, 1861 (1997)
  [hep-ph/9606387].
}



\bibitem{GW} 
  L.~Senatore, E.~Silverstein and M.~Zaldarriaga,
  ``New Sources of Gravitational Waves during Inflation,''
  JCAP {\bf 1408} (2014) 016
  [arXiv:1109.0542 [hep-th]].

\bibitem{Sorbo} 
  L.~Sorbo,
  ``Parity violation in the Cosmic Microwave Background from a pseudoscalar inflaton,''
  JCAP {\bf 1106}, 003 (2011)
  [arXiv:1101.1525 [astro-ph.CO]].

J.~L.~Cook and L.~Sorbo,
  ``Particle production during inflation and gravitational waves detectable by ground-based interferometers,''
  Phys.\ Rev.\ D {\bf 85}, 023534 (2012)
  [Erratum-ibid.\ D {\bf 86}, 069901 (2012)]
  [arXiv:1109.0022 [astro-ph.CO]].


\bibitem{Mukohyama} 
  N.~Barnaby, J.~Moxon, R.~Namba, M.~Peloso, G.~Shiu and P.~Zhou,
  ``Gravity waves and non-Gaussian features from particle production in a sector gravitationally coupled to the inflaton,''
  Phys.\ Rev.\ D {\bf 86}, 103508 (2012)
  [arXiv:1206.6117 [astro-ph.CO]].

  S.~Mukohyama, R.~Namba, M.~Peloso and G.~Shiu,
  ``Blue Tensor Spectrum from Particle Production during inflation,''
  arXiv:1405.0346 [astro-ph.CO].




\bibitem{Behbahani:2011it}
  S.~R.~Behbahani, A.~Dymarsky, M.~Mirbabayi and L.~Senatore,
  ``(Small) Resonant non-Gaussianities: Signatures of a Discrete Shift Symmetry in the Effective Field Theory of Inflation,''
  JCAP {\bf 1212} (2012) 036
  [arXiv:1111.3373 [hep-th]].

\bibitem{Silverstein:2008sg}
  E.~Silverstein and A.~Westphal,
  ``Monodromy in the CMB: Gravity Waves and String Inflation,''
  Phys.\ Rev.\ D {\bf 78} (2008) 106003
  [arXiv:0803.3085 [hep-th]];

L.~McAllister, E.~Silverstein and A.~Westphal,
  ``Gravity Waves and Linear Inflation from Axion Monodromy,''
  Phys.\ Rev.\ D {\bf 82}, 046003 (2010)
  [arXiv:0808.0706 [hep-th]].

R.~Flauger, L.~McAllister, E.~Pajer, A.~Westphal and G.~Xu,
  ``Oscillations in the CMB from Axion Monodromy Inflation,''
  JCAP {\bf 1006}, 009 (2010)
  [arXiv:0907.2916 [hep-th]].

\bibitem{Trapped}

D.~Green, B.~Horn, L.~Senatore and E.~Silverstein,
  ``Trapped Inflation,''
  Phys.\ Rev.\ D {\bf 80}, 063533 (2009)
  [arXiv:0902.1006 [hep-th]].


\bibitem{Unwind} 
  G.~D'Amico, R.~Gobbetti, M.~Kleban and M.~Schillo,
  ``Unwinding Inflation,''
  JCAP {\bf 1303}, 004 (2013)
  [arXiv:1211.4589 [hep-th]].

\bibitem{Barnaby} 
  N.~Barnaby and M.~Peloso,
  ``Large Nongaussianity in Axion Inflation,''
  Phys.\ Rev.\ Lett.\  {\bf 106}, 181301 (2011)
  [arXiv:1011.1500 [hep-ph]].


\bibitem{EFTofI} 
  C.~Cheung, P.~Creminelli, A.~L.~Fitzpatrick, J.~Kaplan and L.~Senatore,
  ``The Effective Field Theory of Inflation,''
  JHEP {\bf 0803}, 014 (2008)
  [arXiv:0709.0293 [hep-th]].


\bibitem{guido}
  G.~D'Amico and M.~Kleban,
  Phys.\ Rev.\ Lett.\  {\bf 113} (2014) 081301
  [arXiv:1404.6478 [astro-ph.CO]].

\bibitem{senatoretalk} 
L.~Senatore: talk at the ``Burke Institute Workshop on PRIMORDIAL GRAVITATIONAL WAVES AND COSMOLOGY'', Caltech, May 2014, slides at https://burkeinstitute.caltech.edu/workshops/BICEP2; also at ICTP Workshop on FRONTIERS OF NEW PHYSICS, COLLIDERS AND BEYOND, June 2014, slides at http://indico.ictp.it/event/a13203/session/3/\\contribution/8/material/2/.


\bibitem{Baumann:2014cja} 
  D.~Baumann, D.~Green and R.~A.~Porto,
  arXiv:1407.2621 [hep-th].

\bibitem{Creminelli} 
  P.~Creminelli, J.~Gleyzes, J.~Noreña and F.~Vernizzi,
  arXiv:1407.8439 [astro-ph.CO].



\bibitem{Goldberger:2004jt}
  W.~D.~Goldberger and I.~Z.~Rothstein,
  ``An Effective field theory of gravity for extended objects,''
  Phys.\ Rev.\ D {\bf 73} (2006) 104029
  [hep-th/0409156].

\bibitem{Weinberg}

S.~Weinberg, Gravitation and Cosmology (Wiley, New York, 1972)


\bibitem{Senatore:2010wk}
  L.~Senatore and M.~Zaldarriaga,
  ``The Effective Field Theory of Multifield Inflation,''
  JHEP {\bf 1204} (2012) 024
  [arXiv:1009.2093 [hep-th]].

\bibitem{Ozsoy} 
  O.~Özsoy, K.~Sinha and S.~Watson,
  arXiv:1410.0016 [hep-th].



\bibitem{Ferreira} 
  R.~Z.~Ferreira and M.~S.~Sloth,
  arXiv:1409.5799 [hep-ph].

\bibitem{Matteo} 
  M.~Biagetti, E.~Dimastrogiovanni, M.~Fasiello and M.~Peloso,
  arXiv:1411.3029 [astro-ph.CO].


\bibitem{Adshead} 
  P.~Adshead, E.~Martinec and M.~Wyman,
  Phys.\ Rev.\ D {\bf 88}, no. 2, 021302 (2013)
  [arXiv:1301.2598 [hep-th]].

\bibitem{Creminelli:2005hu}
  P.~Creminelli, A.~Nicolis, L.~Senatore, M.~Tegmark and M.~Zaldarriaga,
  ``Limits on non-gaussianities from wmap data,''
  JCAP {\bf 0605} (2006) 004
  [astro-ph/0509029].


\bibitem{Senatore:2009gt}
  L.~Senatore, K.~M.~Smith and M.~Zaldarriaga,
  ``Non-Gaussianities in Single Field Inflation and their Optimal Limits from the WMAP 5-year Data,''
  JCAP {\bf 1001} (2010) 028
  [arXiv:0905.3746 [astro-ph.CO]].

\bibitem{Klein} 
  S.~Klein,
  ``Suppression of Bremsstrahlung and pair production due to environmental factors,''
  Rev.\ Mod.\ Phys.\  {\bf 71}, 1501 (1999)
  [hep-ph/9802442].


\bibitem{Koshelkin} 
  A.~V.~Koshelkin,
  ``Long wave asymptote for the Landau-Pomeranchuk-Migdal effect,''
  J.\ Phys.\ A {\bf 37}, 1051 (2004).


\bibitem{Sasaki} 
  M.~Sasaki and E.~D.~Stewart,
  ``A General analytic formula for the spectral index of the density perturbations produced during inflation,''
  Prog.\ Theor.\ Phys.\  {\bf 95}, 71 (1996)
  [astro-ph/9507001].




\end{thebibliography}
\end{document}